# HYDROGRAPHIC VARIABILITY AND BIOMASS FLUCTUATIONS OF EUROPEAN ANCHOVY (ENGRAULIS ENCRASICOLUS) IN THE CENTRAL MEDITERRANEAN SEA: MONETARY ESTIMATIONS AND IMPACTS ON FISHERY FROM LAGRANGIAN ANALYSIS

Antonio Di Cintio [1,2], Marco Torri [3], Federico Falcini [1], Raffaele Corrado [1], Guglielmo Lacorata [1,4], Angela Cuttitta [3], Bernardo Patti [3], Rosalia Santoleri [1]

[1] Institute of Marine Sciences, National Research Council (CNR), Rome, Italy

[2] Anton Dohrn Zoological Station, Napoli, Italy

[3] Institute of Anthropic Impacts and Sustainability in marine environment (IAS), National Research Council (CNR), detached unit of Capo Granitola, Campobello di Mazara, Italy.

[4] Center of Excellence for Telesensing of Environment and Model Prediction of Severe events (CETEMPS), Università dell'Aquila, L'Aquila, Italy.

**Abstract**

During the last decades, scientific community has been investigating both biological and hydrographic processes that affect fisheries. Such an interdisciplinary and synergic approach is nowadays giving a fundamental contribution, in particular, in connecting the dots between hydrographic phenomena and biomass variability and distribution of small pelagic fish. Here we estimate impacts of hydrographic fluctuations on small pelagic fishery, focusing on the inter-annual variability that characterizes connectivity between spawning and recruiting areas for the European anchovy (*Engraulis encrasicolus*, Linnaues 1758), in the Northern side of the Sicily Channel (Mediterranean Sea). Results show that coastal transport dynamics of a specific year largely affect the biomass recorded the following year. Our work, moreover, quantifies the specific monetary impacts on landings of European anchovy fishery due to hydrodynamics variability, connecting biomass fluctuations with fishery economics in a highly dynamic and exploited marine environment as the Sicily Channel. In particular, we build a model that attributes a monetary value to the hydrographic phenomena (i.e., cross-shore vs. alongshore eggs and larvae transport), registered in the FAO Geographical Sub-Area (GSA) 16 (Southern Sicily). This allows us to provide a monetary estimation of catches, derived from different transport dynamics. Our results highlight the paramount importance that hydrographic phenomena can have over the socio-economic performance of a fishery.

1. **Introduction**

Small pelagic fish play a key role in the food web of marine pelagic ecosystem, representing a connection between lower trophic levels and top predators (Cury et al., 2000; Bănaru et al., 2013). The European anchovy (*Engraulis encrasicolus*, Linnaues 1758) finds suitable spawning grounds across the whole Mediterranean Sea (Olivar et al., 2001; Zarrad et al., 2006; Somarakis and Nikolioudakis, 2007; Cuttitta et al., 2016a), and constitutes one of the most important species in terms of biomass in the basin (Lleonart & Maynou, 2003, FAO-GFCM 2015). Therefore, biomass fluctuations in European anchovy populations can generate considerable consequences on both the ecosystem structure and functioning (e.g., Cury et al., 2000; Shannon et al., 2000; Daskalov, 2002).

Typically, biomass fluctuations can be associated to the inter-annual variability of environmental parameters (e.g., sea surface temperature and chlorophyll), which play a role in the recruitment of new generations of adult populations by tuning mortality rates in the early life stage (Hunter and Alheit, 1995; Bakun, 1996; Patti et al. 2004; Cuttitta et al., 2015, 2016b; Falcini et al., 2015; Torri et al., 2018). These dynamics are particularly effective in the population of European anchovy, whose short lifespan and large fecundity, coupled with a significant fishing pressure, imply that the age structure of the population is composed essentially by individuals of a few age-classes (Basilone et al., 2000). Therefore, in the framework of fishery management, it is essential determining the relationship between year-to-year recruitment success and stock biomass of exploited species.

Economically, European anchovy and European pilchard (*Sardina pilchardus*) represent more than one third of Italian marine catches. Despite the latter species represents the most caught small pelagic species across all Mediterranean fisheries, anchovies are generally characterized by a higher market price (Quinci, 2011; Patti et al., 2009). As such, these two species are of paramount socio-economic importance for both the Italian fishing and processing sectors.

Sciacca, in particular, is the main small pelagic fishing port in Southern Sicily (Fig. 1). In this port, catches of sardines and anchovies jointly represent more than 90% of small pelagic landings (Mazzola et al., 2002, Patti et al., 2004). Fishing is implemented through pelagic pair trawling and lamplight purse-seining. Pelagic trawlers land all their catch in Sciacca (Patti et al., 2010). While a small portion of the catch is destined to direct human consumption, the vast majority of landings are sent to processing plants (Patti et al., 2014).

By playing a predominant role in the recruitment success of European anchovy stock in the FAO Geographical Sub-Area (GSA) 16 (Southern Sicily), environmental patterns and ocean dynamics significantly influence adult population biomass (Quinci, 2011; Palatella et al., 2014), as well as the spatial distribution of early life stage individuals (Cuttitta et al., 2016b, 2018; Patti et al., 2018). Several investigations demonstrated that while the meandering pattern of the Atlantic Ionian Stream (AIS) plays a fundamental role during the spawning activity, wind-induced coastal currents ensure the connectivity between spawning and recruiting areas (García Lafuente et al., 2005; Palatella et al., 2014; Falcini et al., 2015). However, too intense wind activity (i.e., Mistral bursts) may give rise to mesoscale structures, such as jets and filaments, which deliver larvae cross-shore (Falcini et al., 2015; Torri et al., 2018). Therefore, the success of the connectivity between spawning grounds and the recruiting zone off Cape Passero (Fig. 1) is strongly regulated by the variability in environmental factors, most specifically in marine currents.

Palatella et al. (2014) investigated the Lagrangian connectivity between an ideal spawning area off Sciacca (with spawning activity generally peaking around July-August) and the recruiting zone off Capo Passero, from 1999 to 2012. They assumed that virtual larvae, crossing the imaginary line linking Malta to Sicily (Fig. 1), generally reach the recruiting zone off Cape Passero and thus survive. On the other hand, those larvae that do not cross this line are transported offshore and die. Results showed a strong and intriguing inter-annual variability in this process (Table 1) that poses the following question: what is the monetary impact of Lagrangian connectivity with respect to a successful, highly productive spawning? Based on the assumptions set by Palatella et al. (2014), this paper aims at assessing the monetary value of climate variability affecting those oceanographic patterns that rule the connectivity between spawning and recruiting areas of European Anchovy in GSA 16.

2. **Data and Methods**

*2.1   Estimation of "survived eggs" parameter, catch weight, and monetary values*

Data on anchovy egg density were collected in GSA 16 (Fig. 1) during yearly oceanographic surveys ANSIC/BANSIC, performed between 1999 and 2012. Surveys were run in the summer season i.e. in correspondence of the spawning period (Cuttitta et al., 2003 and 2015). Samplings were carried out on a regular grid of stations (1/10° × 1/10° along the continental shelf) (Fig. 1). In each point, oblique tows of Bongo40 net, composed by two coupled mouths of 40 cm and equipped with 200 μm mesh size net, were carried out with the aim to collect mesozooplanktonic samples. Biological samples were preserved in 70% ethanol solution. Eggs were identified and counted by using a stereomicroscope. The filtered water volume of each mouth was measured by a calibrated flow-meter (G.O. 2030). An estimation of the density of anchovy eggs, i.e., $\rho_{eggs}$ (number/m$^3$), was performed in each sampling station. The total amount of eggs estimated to be found within the spawning area off Sciacca (Fig. 1) during the summer spawning of each *i*-th year, is:

$$Tot\_eggs_i = \bar{\rho}_{eggs_i} * A * h, \tag{1}$$

where, $\bar{\rho}_{eggs_i}$ is the average egg density, calculated for each *i*-th year during the summer-spawning surveys, along the whole coastal area of GSA 16 (Fig. 1); $A$ is the size of the spawning area; and $h \sim 10\ m$ is a vertical scale value, chosen by considering that anchovy's eggs in the Mediterranean Sea generally float around the sea surface, with 95.5% of them found within the upper 10 meters of the water column (Palomera, 1991). Anchovy eggs also reach depths of 50 meters, despite very few of them achieve to do so (Ospina-Álvarez et al. 2012). Therefore, we followed a conservative approach, based on the first 10 meters of the water column.

We are aware that $Tot\_eggs_i$ in Equation (1) is an underestimation of the total number of eggs realistically released during the whole spawning period by the European anchovy in the GSA 16. Indeed, our surveys allow obtaining a good estimation of this value in correspondence of the survey period (two to three weeks between June and August), i.e. during the spawning peak. However, the spawning period of the European anchovy in the Strait of Sicily occurs from May to November, with a peak in the warmest period of the summer season. It follows that our estimation of the total eggs is necessarily lower due to the lack of the integration of the estimate over the whole spawning period. Given the extreme variability of the duration and indensity of the spawning period, we avoid the application of a multiplicative factor, indicative of the integration over time, preferring an analysis of the relative variations and trends during the years considered in the present study.

To represent the transport dynamics of eggs and larvae within GSA 16, i.e. the Lagrangian connectivity between spawning grounds and the recruiting area off Capo Passero, we consider the virtual spawning area off Sciacca (Fig. 1) (Palatella et al., 2014). From this area, Lagrangian trajectories that statistically represent the path of larvae allow us quantifying the rate of connectivity with Capo Passero by means of a Lagrangian Transport Index (LTI). Such an index is thought to estimate the efficiency of the coastal current (Fig. 1) in connecting spawning to nursery areas. This indicator is the integral of the probability distribution function over the arrival times of eggs and larvae that travel from the spawning area near Sciacca to the nursery area off Cape Passero. Arrivals are evaluated when individuals cross the imaginary line between Malta and Sicily (Fig. 1). More specifically, the LTI is the percentage of those individuals that arrive to the recruitment area within 25 days after release, taking into consideration the whole reproductive period (Palatella et al., 2014).

Indeed, only a given percentage of all those larvae that are produced off Sciacca and, in general, along the whole spawning area, is expected to reach the adult phase. Most of the recruitment success will depend on whether the larvae will be able to reach the nursery area off Cape Passero. The LTI, therefore, quantifies the fate of these larvae and thus the efficiency of coastal currents in carrying anchovy eggs and larvae off Cape Passero. Hence, we multiply the total eggs that are estimated along the coastal spawning zone by in situ samplings (i.e., Equation 1) by the LTI, thus providing for each year ($i$) the percentage of those larvae that are expected to reach Cape Passero, i.e., *Survived$_i$*:

$$Survived_i = Tot\_eggs_i * LTI_i, \qquad (2)$$

where we recall that the LTI is a non-parametric, dimensionless index, thus applicable to the whole coastal area of the GSA 16 (Fig. 1). Accordingly, we consider "not-survived individuals" those larvae, post-larvae, and eggs that are spread across a 25-days scale and did not cross the imaginary line between Malta and Sicily (Fig. 1), thus not becoming adults.

To evaluate the impact of a successful alongshore transport over the European anchovy fishery in GSA 16, we provide a monetary estimation to the amount of eggs and larvae that manage to reach Cape Passero and survive. To achieve this goal, we first notice that not all the fish that compose a given stock do end up in fishing nets, since only a certain percentage is actually fished, depending on the stock exploitation rate. We therefore estimate the exploitation rate of the European anchovy fishery in GSA 16 by considering the ratio between catches and biomass, i.e., $\frac{catches_i}{biomass_i}$, recorded in the same year. Anchovy biomass and landing data for the whole GSA 16 are provided by GFCM (2012). We estimated catches from landing data by considering a 5% discard rate taking place in pelagic trawl and purse seine fisheries targeting small pelagics (mainly anchovies) in Eastern and Central Mediterranean Sea (Kallianiotis and Mazzola 2002). In GSA 16, purse seiners and pelagic trawlers targeting small pelagics are mainly based in the port of Sciacca, accounting for about 2/3 of total GSA 16 landings (GFCM 2012).

To calculate the amount of eggs and larvae that each year is converted into "fishable" individuals, we consider the total amount of fish caught in that year ($Tot\_catch_i$) to be the sum of two catch components, i.e., $Catch1_i$ and $Catch2_i$. The former is the amount of eggs that were laid the previous year and survived ($Survived_{i-1}$), multiplied by the exploitation rate $\frac{catches_i}{biomass_i}$ of the fishery in *i*-th year. Basilone et al. (2000) report that 95% of anchovy catches in the Sicily Channel is composed of individuals of age 1 and 2 years-old while individuals of three or more years account for a maximum of 5% of the catch. Therefore, we consider only a two-year interval for assessing how survived eggs will contribute to future catches, i.e., all the fish is considered to be caught either in its first or second year of life. It follows that $Catch\ 2_i$ is composed by the amount of eggs that were laid two years before, and then survived ($Survived_{i-2}$), multiplied by the exploitation rate of the current year $\left(\frac{catches_i}{biomass_i}\right)$, and by the survival rate of individuals of the stock in year *i*, represented by the parameter $1 - \frac{catches_{i-1}}{biomass_{i-1}}$. We assume that all the fish that is not fished today, survives until the next fishing season, i.e., factors such as natural mortality or predation from other species are not considered. It results, therefore, that:

$$Tot\_catch_i = Catch1_i + Catch2_i, \qquad (3)$$

Where:

$$\text{Catch1}_i = Survived_{i-1} \left(\frac{catches_i}{biomass_i}\right) \tag{4}$$

And:

$$\text{Catch2}_i = Survived_{i-2} \left(1 - \frac{catches_{i-1}}{biomass_{i-1}}\right)\left(\frac{catches_i}{biomass_i}\right). \tag{5}$$

We remind that $Survived_i$ is defined in Equation (2) and, therefore, equation (3) includes the spawning-recruiting connectivity factor given by the LTI. Moreover, we are aware of the fact that the exploitation rates we use in equations (4) and (5) are related to the whole stock, and thus, to all areas of GSA 16.

Each anchovy adult weights ~13.2 g (average weight obtained by 18 hauls carried out during the acoustic survey ANCHEVA07/00 on the continental shelf off the southern coast of Sicily in July 2000, and reported in Mazzola et al. 2002). From this, we can estimate the total weight of the catch, i.e., $Weight\_LTI_i$ (Table 2) as:

$$Weight\_LTI_i = Tot\_catch_i * 0.0132. \tag{6}$$

The mean adult weight we consider is associated to a maturity size of 9.7 cm (average size ranging from 9 to 14 cm; Whitehead et al., 1988). However, it is worth noting that, due to overfishing occurring in the anchovy fishery in the Sicilian Channel (GFCM 2012), it is possible that the average size and weight of adult individuals has decreased with respect to the reported values.

To provide a monetary estimation of the variability in the amount of larvae that are transported by coastal currents to the recruiting area, we consider the mean price per kilogram ($Avg\_price_i$) of anchovy, recorded every year in the market of Aci Trezza (Catania, Sicily), as provided by the Italian Institute for Services for the Agricultural and Food Market (ISMEA 2017). By multiplying this value by the total weight of the catch, we obtain the monetary value of catches, for each year, that can be associated to the marine current dynamics whose variability affects eggs and larvae survival (Table 2):

$$Value\_LTI_i = Weight\_LTI_i * Avg\_price_i * CPI_i, \qquad (7)$$

where we consider the decimal form of the Italian consumer price index ($CPI_i$), provided by ISTAT (2017), which is re-indexed to the year 2001, in order to compare the real monetary value of catches across different years.

We finally remark that our $Value\_LTI_i$ is not meant to represent the actual market value of catches, but rather a monetary estimation of how the environmental variability affects stock biomass.

## 2.2 *Cross-Correlation among the key variables and the regression model selection*

We calculate the Pearson correlation coefficients and relative p-values for anchovy biomass and the three following variables, evaluated for one (*i-1*) year (Table 3): i) the total amount of produced eggs (i.e., $Tot\_eggs$ in Equation 1), ii) the $LTI$, and iii) the amount of larvae, survived taking into account the total amount of eggs and the LTI, i.e., $Survived$ in Equation (2). Correlations are evaluated by excluding the year 2000 (Table 3), since it shows non-reliable data: the expected correlation between biomass and laid eggs is indeed not realistic (or, at least, affected by large error) when compared to the other years (Table 1). This is likely due to the fact that the coverage of the investigated area for the ichthyoplanktonic survey carried out in 1999 was significantly lower than the other years (Mazzola et al., 2002).

The relation between the response variable *biomass$_i$* and the explanatory variables $Tot\_eggs_i$, $LTI_i$, $Survived_i$, lagged by one year, is investigated by means of a multivariate analysis. Regression models were implemented in order to explore how variables relate, and to identify the most important factors that affect the response variable. In order to deal with the possible issue of multi-collinearity prior to model selection (Neter et al., 1996), the Pearson product-moment correlation coefficient was considered before the implementation of the model (e.g. among $Tot\_eggs_i$ or $LTI_i$ and $Survived_i$ that is a derived parameter) and, eventually, correlated variables were considered in separated models. Normality distribution of the response variable $Biomass_i$ is tested by using Shapiro-Wilk normality test. Since the response variable is normally distributed (p-value > 0.05), we consider a Gaussian distribution in all models and an "identity" link between the mean of the response variable and the systematic part of the model. Linearity and non-linear relationships are examined by means of Generalized Linear Models (GLM) and Generalized Additive Models (GAM) (Wood, 2011), respectively. Akaike's Information Criterion (AIC) (Akaike, 1973) is used to compare models and determine which model(s) served as the best approximation(s) to the data. For the validation model, normal Q-Q plot of the residuals, plot of residuals vs. fitted values, and plot of residuals vs. variables considered in the model are used to assess normality distribution of the residuals and the assumptions of homogeneity and independence of the variables (Zuur et al., 2009). All analysis were performed with R software version 3.5.0. (R Core Team, 2018). All statistical models are implemented using "mgcv" R package (Wood, 2011).

### 3. Results

Correlation coefficients as well as the regression models resulting from the comparison among key variables (Tables 3 and 4, Fig. 2) highlight the key role of total egg production ($Tot\_eggs_i$) and surface physical forcing ($LTI_i$) in driving the fluctuations of European anchovy biomass in GSA 16. Table 3 shows the correlation among the total amount of produced eggs, the $LTI_i$, the amount of larvae that survived and reached Cape Passero, and the level of biomass recorded the following year. Correlation between egg production and the following year's biomass is not significant: a large amount of eggs does not seem to affect the following year's biomass levels (-0.07). Moreover, p-value for this correlation is 0.77, which indicates that the relation is not statistically significant. On the other hand, $LTI_i$ has some tangible implications in determining the amount of biomass experienced the following year. Indeed, the correlation between $LTI_i$ and next-year (i.e., *i*+1) biomass is positive (i.e., 0.41), despite a p-value of 0.20. The amount of survived eggs (i.e., equation 2) showed the highest correlation coefficients (0.59), with a p-value of 0.05. Thus, when considered jointly, the number of eggs as well as the oceanographic conditions occurring during spawning are crucial in affecting the biomass of anchovy population in the following year.

Regression models confirm these findings and provide a more efficient description of the relationship among the key variables (Table 4 and Fig. 2). Two separated models were implemented in order to avoid a multi-collinearity issue: one considering $Tot\_eggs_i$ and $LTI_i$, and one considering only $Survived_i$ parameters as explanatory variable. GLM did not show any significant term. On the other hand, GAM highlighted in each case significant cubic regression splines. In particular, the best model results to be the GAM with the $Survived_i$ parameter (Tab. 4), suggesting the importance of both parameters (i.e., $Tot\_eggs_i$ and $LTI_i$,) in influencing the biomass fluctuations occurring the following year. Moreover, the shape of the cubic regression spline (Fig. 2) evidences a non-linear relationship and suggests a strong, positive correlation when $Survived_i$ reaches a threshold of 15 million eggs. We remark, however, the underestimation of the total eggs in terms of absolute value due to the features of the sampling design.

A non-negligible evidence emerges also from the analysis of the other GAM model, i.e., $Biomass_i \approx s(Tot\_eggs_{i-1}) + s(LTI_{i-1})$. The significance of the LTI, as well as the almost identical trend of the $LTI$ and $Survived$ splines (Fig. 2), suggest that the relationship between biomass and the eggs survived from the previous year are strictly linked to the LTI parameters. That is, biomass is strongly affected by the number of eggs that are advected from the main spawning area (in front of Sciacca) to the recruitment area of Capo Passero.

Table 2 reports the estimation (weight and value of the catch) of the impact of alongshore transport on the European anchovy fishery in GSA 16, as obtained from equations (3)-(7). We observe a very high inter-annual variability in both catch weight and value (the latter two variables are strongly positively correlated). The former variable ranges from a minimum of 228 kg in 2002, to a maximum of 486 625 kg in 2006 (Table 2). Likewise, real value of the catch ranges from 856 € in 2002 to 2 450 394 € in 2006 (Table 2). The high variability is due to the changes in hydrographic phenomena (i.e., LTI) as well as in the exploitation rates calculated across the years from official biomass and catch data, as shown by equations (3)-(5).

4. Discussions and conclusions

To evaluate the impacts of environmental variability on the European anchovy fishery in GSA 16, we assessed the influence of inter-annual variability of hydrographic phenomena in the Sicilian Channel on egg and larval dispersion and biomass. This variability is mainly represented by changes in the transport properties of the coastal current that delivers anchovy eggs and larvae alongshore, from the spawning ground of Sciacca to the recruitment area off Cape Passero.

We showed how the success of alongshore transport is crucial for ensuring adequate recruitment levels for anchovy larvae. The amount of larvae reaching Cape Passero and originating from the spawning areas located upstream is actually correlated to the biomass, recorded in the following years. That is, a higher European anchovy biomass in the Mediterranean GSA 16 is recorded when a favourable combination of advective currents and egg deposition is observed in the previous year. Since, from Equation (2), $Survived_i = Tot\_eggs_i * LTI_i$, the combination of both variables $Tot\_eggs$ and $LTI$ is what mainly drives the biomass. On the other hand, a large amount of egg deposition does not itself appear to be crucial in affecting the following year's biomass. The non-significant statistical correlation strengthens the hypothesis that the total amount of laid eggs is not the main parameter that should be taken into account when trying to identify those key factors that affect the following year's biomass availability.

We do not calculate the Pearson correlation coefficients between $Tot\_eggs_i$, $Survived_i$, and $LTI_i$, and $Weight\_LTI_i$. This is because the latter variable is determined by the exploitation rates calculated each year for GSA 16, which are not related to the three former variables composing the monetary estimation we present in our study. Therefore, we do not expect any kind of correlation among these variables (additionally, no correlation takes place between $Tot\_eggs_i$, $Survived_i$ $LTI_i$, and the official landing data reported by GFCM 2012). Likewise, we do not perform a Pearson correlation test between the three variables we mentioned above and $Value\_LTI_i$. Moreover, comparing those three variables with the monetary value of the catch would be misleading, due to the fact that the latter is based on average per kilo prices that are not related to $Tot\_eggs_i$, $Survived_i$ and $LTI_i$. Additionally, average prices are also influenced by phenomena occurring outside GSA 16 (e.g. competing neighbouring fisheries). Lastly, average prices can be affected by the economic performance of a country as well as its inflation rate. For instance, consider the years 2010 and 2011: the sharp decrease in the CPI caused the difference in the value of the catch between 2010 and 2011 to rise from around 39 thousand € (in nominal terms) to around 74 thousand € (in real terms).

By using Equations (3)-(7), we provide a monetary estimation of the impact of alongshore transportation patterns on fish catches. That is, we give a monetary value to the amount of eggs and larvae that manage to reach Cape Passero and survive. This allows providing a quantifiable value to the hydrographic phenomena registered in GSA 16 (Table 2). This estimation aims at quantifying the potential impacts that marine currents can have on one of the main aspects affecting fisheries socio-economics, i.e., the gross revenues deriving from fishing operations. We observe a significant inter-annual variability among the value of the catch that we estimate (Table 2).

The evolution of prices and inflation rates affects the estimate value of the catch (Table 2). The "valuation" of the impacts of hydrographic phenomena over a fishery changes from one year to the other, according to the different circumstances that take place both inside and outside the fishery into account. We observe, in particular, significant differences between the maximum and minimum annual anchovy prices in the market of Aci Trezza (Table 2). One possible explanation could be linked to the seasonality of fishing operations. This implies that the value of the catch could rise or decrease significantly, according to the time of the year in which it is sold. Aci Trezza, indeed, is a production market, where goods are offered directly by single producers or producers associated into cooperatives. Additionally, the value of other activities that are related to fishery (e.g., processing, distribution, retailing, etc…), as well as the fiscal revenue they generate, is not considered in the present paper. This means that the "value" of hydrographic phenomena provided in this paper might be underestimated.

Concluding, in our work we shed some light over the impacts that marine currents can have over biomass fluctuations of European Anchovy in GSA 16. Similar investigations could be applied to several other fisheries. Considering only biological parameters such as the spawning stock biomass or the amount of laid eggs, while ignoring oceanographic factors such as alongshore transportation, can lead to biased biomass expectations as well as management decisions.

**Table 1.** Anchovy egg density (data provided by the Italian National Research Council), total egg production (obtained by multiplying the average egg density by the area of the study region from Palatella et al. (2014) (783 000 000 m$^2$) and by 10, LTI (the percentage of individuals arriving alive in the recruitment area within 25 days after release; Palatella et al. 2014), survived eggs, biomass (GFCM 2012), landings (GFCM 2012), catches and exploitation rate.

| Year | Avg_egg (n° of eggs/m³) | Tot_eggs (n° of eggs laid in the study area) | LTI | Survived_eggs (n. of survived individuals) | Anchovy biomass (tons) | Anchovy landings (tons) | Anchovy catches (tons) | Exploitation rate |
|---|---|---|---|---|---|---|---|---|
| 1999 | 0.116623 | 913 505 155 | 0.000889 | 812 106 | 20 200 | 2 043 | 2 145 | 0.106196 |
| 2000 | 0.168873 | 1 322 782 334 | 0.000003 | 3 968 | 11 000 | 189 | 198 | 0.018041 |
| 2001 | 0.218891 | 1 714 570 266 | 0.000033 | 56 581 | 22 950 | 1 627 | 1 708 | 0.074438 |
| 2002 | 0.143400 | 1 123 252 200 | 0.005994 | 6 732 774 | 11 500 | 3 294 | 3 459 | 0.300757 |
| 2003 | 0.096835 | 758 512 025 | 0.000094 | 71 300 | 9 200 | 2 218 | 2 329 | 0.253141 |
| 2004 | 0.514786 | 4 032 316 498 | 0.012556 | 50 629 766 | 9 820 | 1 554 | 1 632 | 0.166161 |
| 2005 | 0.355512 | 2 784 727 023 | 0.003704 | 10 314 629 | 20 702 | 2 390 | 2 510 | 0.121220 |
| 2006 | 0.406475 | 3 183 921 887 | 0.002828 | 9 004 131 | 6 370 | 4 262 | 4 475 | 0.702527 |
| 2007 | 0.275795 | 2 160 301 232 | 0.003444 | 7 440 077 | 6 725 | 4 812 | 5 053 | 0.751316 |
| 2008 | 0.137197 | 1 074 666 984 | 0.006492 | 6 976 738 | 3 130 | 1 062 | 1 115 | 0.356262 |
| 2009 | 0.192961 | 1 511 465 401 | 0.001301 | 1 966 417 | 5 833 | 4 302 | 4 517 | 0.774404 |
| 2010 | 0.523749 | 4 102 529 301 | 0.000343 | 1 407 168 | 15 880 | 5 124 | 5 380 | 0.338804 |

| 2011 | 0.199523 | 1 562 862 860 | 0.001004 | 1 569 114 | 5 092 | 4 018 | 4 219 | 0.828535 |
| 2012 | 0.085949 | 673 239 238 | 0.010199 | 6 866 367 | 10 419 | 2 625 | 2 756 | 0.264541 |

**Table 2.** Anchovy catch composition, weight, average price in Aci Trezza market (ISMEA undated), and nominal and real value of catch in GSA 16 (as a consequence of LTI, total egg production, exploitation rate, prices in Aci Trezza market and Italian inflation rate).

| Year | Catch_1 (n° of individuals) | Catch_2 (n° of individuals) | Total_catch (n° of individuals) | Weight_LTI (Total weight of the catch, kg) | Avg_€_AT (Average price and standard deviation in Aci Trezza, €/kg) | Nominal value of catch | CPI (1995=100) | Reindexed_CPI (Decimal form of CPI reindexed to 2001) | Value_LTI (Real value of catch in 2001 terms) |
|---|---|---|---|---|---|---|---|---|---|
| 2001 | 295 | 59 361 | 59 656 | 787 | 4.45 (2.14) | 3 508 € | 115.1 | 1 | 3 508 € |
| 2002 | 17 017 | 1 105 | 18 122 | 239 | 3.86 (2.38) | 923 € | 117.9 | 1.028 | 898 € |
| 2003 | 1 704 343 | 10 015 | 1 714 358 | 22 630 | 5.05 (4.60) | 114 279 € | 120.8 | 1.057 | 108 116 € |
| 2004 | 11 847 | 835 529 | 847 376 | 11 185 | 3.75 (2.38) | 41 945 € | 123.2 | 1.081 | 38 802 € |
| 2005 | 6 137 349 | 7 207 | 6 144 556 | 81 108 | 4.98 (3.17) | 403 919 € | 125.3 | 1.102 | 366 532 € |
| 2006 | 7 246 310 | 31 257 145 | 38 503 455 | 508 246 | 5.67 (4.49) | 2 884 294 € | 127.8 | 1.127 | 2 559 267 € |
| 2007 | 6 764 948 | 2 305 277 | 9 070 225 | 119 727 | 5.17 (3.78) | 618 988 € | 130.0 | 1.149 | 538 719 € |
| 2008 | 2 650 617 | 797 736 | 3 448 353 | 45 518 | 5 (3.54) | 227 591 € | 134.2 | 1.191 | 191 093 € |
| 2009 | 5 402 816 | 3 708 979 | 9 111 794 | 120 276 | 4.83 (2.65) | 581 533 € | 135.2 | 1.201 | 484 207 € |

| | 2010 | 666 229 | 533 250 | 1 199 479 | 15 833 | 5.68 (2.38) | 89 932 € | 137.3 | 1.222 | 73 594 € |
|---|---|---|---|---|---|---|---|---|---|---|
| | 2011 | 1 165 888 | 1 077 251 | 2 243 138 | 29 609 | 4.37 (1.63) | 129 393 € | 102.7 | 0.876 | 147 709 € |

**Table 3.** Pearson correlation coefficients between total egg production, LTI, survived eggs, and the following years' biomass. Values in parenthesis indicate the p-value (i.e., Pearson correlation coefficient).

| Time series | | Total eggs$_i$ | LTI$_i$ | Survived$_i$ |
|---|---|---|---|---|
| 2001-2011 | Biomass$_{i+1}$ | -0.07 (0.77) | 0.41 (0.20) | 0.59 (0.05) |

Table 4. GLM and GAM models implemented considering the biomass (year *i*) and the monetary values of the catches (year *i*) as dependent variables and the parameters total eggs production (years *i+1*), LTI (years *i+1*) and the survived eggs(years *i+1*) as independent variables. Prior to the implementation of the models, the multicollinearity among independent variables was checked and separated models were carried out in the positive cases. In red has been evidenced the best models based on the Akaike's Information Criterion, AIC.

| Model | Time series (i) | Formula | sign terms (p value) | r2 (adj) | Dev. Expl. | AIC |
|---|---|---|---|---|---|---|
| GLM | 2001-2011 | $biomass_i \approx tot\_egg_{i-1} + LTI_{i-1}$ | no sign. | | | |
| GLM | 2001-2011 | $biomass_i \approx survived_{i-1}$ | no sign. | | | |
| GAM | 2001-2011 | $biomass_i \approx s(tot\_egg_{i-1}) + s(LTI_{i-1})$ | $LTI_{i-1}$ (Fig. 2c) | 0.049 | 0.401 | 51.7% | 224.5618 |
| GAM | 2001-2011 | $biomass_i \approx s(survived_{i-1})$ | $survived_{i-1}$ (Fig. 2d) | 0.0493 | 0.421 | 52.0% | 223.3429 |

**Figure Captions**

**Figure 1.** Sicilian Channel bathymetry, anchovy main spawning area (orange area, measuring approximately 783.3 km$^2$), and line of demarcation marking the target area in the Lagrangian numerical simulation. Source: Palatella et al. (2014).

**Figure 2.** Significant cubic regression splines implemented in the GAM models.